\documentclass{webofc}
\usepackage{hhline}
\usepackage{aas_macros}
\usepackage{subcaption}
\usepackage{color}
\usepackage[varg]{txfonts}   % Web of Conferences font
\newcommand{\om}{\Omega_\textrm{m}}
\newcommand{\sig}{\sigma_\textrm{8}}

\newcommand{\planck}{{\em Planck}}

\begin{document}
\title{PSZ2 G091 :  A massive double cluster at $z\sim 0.822$ observed by the NIKA2 camera}

\author {\firstname{E.}~\lastname{Artis}\inst{1}\fnsep\thanks{\email{emmanuel.artis@lpsc.in2p3.fr}\inst{\ref{LPSC}}}
   \and \firstname{R.}~\lastname{Adam} \inst{\ref{LLR}}
   \and  \firstname{P.}~\lastname{Ade} \inst{\ref{Cardiff}}
    \and  \firstname{H.}~\lastname{Ajeddig} \inst{\ref{CEA}}
   \and  \firstname{P.}~\lastname{Andr\'e} \inst{\ref{CEA}}
   \and \firstname{M.}~\lastname{Arnaud} \inst{\ref{CEA}}
   \and  \firstname{H.}~\lastname{Aussel} \inst{\ref{CEA}}
   \and  \firstname{I.}~\lastname{Bartalucci} \inst{\ref{Milano}}
   \and  \firstname{A.}~\lastname{Beelen} \inst{\ref{IAS}}
   \and  \firstname{A.}~\lastname{Beno\^it} \inst{\ref{Neel}}
   \and  \firstname{S.}~\lastname{Berta} \inst{\ref{IRAMF}}
   \and  \firstname{L.}~\lastname{Bing} \inst{\ref{LAM}}
   \and  \firstname{O.}~\lastname{Bourrion} \inst{\ref{LPSC}}
   \and  \firstname{M.}~\lastname{Calvo} \inst{\ref{Neel}}
   \and  \firstname{A.}~\lastname{Catalano} \inst{\ref{LPSC}}
   \and  \firstname{M.}~\lastname{De~Petris} \inst{\ref{Roma}}
   \and  \firstname{F.-X.}~\lastname{D\'esert} \inst{\ref{IPAG}}
   \and  \firstname{S.}~\lastname{Doyle} \inst{\ref{Cardiff}}
   \and  \firstname{E.~F.~C.}~\lastname{Driessen} \inst{\ref{IRAMF}}
   \and  \firstname{A.}~\lastname{Ferragamo} \inst{\ref{Roma}}
   \and  \firstname{A.}~\lastname{Gomez} \inst{\ref{CAB}}
   \and  \firstname{J.}~\lastname{Goupy} \inst{\ref{Neel}}
   \and  \firstname{F.}~\lastname{K\'eruzor\'e} \inst{\ref{LPSC}}
   \and  \firstname{C.}~\lastname{Kramer} \inst{\ref{IRAME}}
   \and  \firstname{B.}~\lastname{Ladjelate} \inst{\ref{IRAME}}
   \and  \firstname{G.}~\lastname{Lagache} \inst{\ref{LAM}}
   \and  \firstname{S.}~\lastname{Leclercq} \inst{\ref{IRAMF}}
   \and  \firstname{J.-F.}~\lastname{Lestrade} \inst{\ref{LERMA}}
   \and  \firstname{J.-F.}~\lastname{Mac\'ias-P\'erez} \inst{\ref{LPSC}}
   \and  \firstname{A.}~\lastname{Maury} \inst{\ref{CEA}}
   \and  \firstname{P.}~\lastname{Mauskopf}
\inst{\ref{Cardiff},\ref{Arizona}}
   \and \firstname{F.}~\lastname{Mayet} \inst{\ref{LPSC}}
   \and  \firstname{A.}~\lastname{Monfardini} \inst{\ref{Neel}}
   \and \firstname{M.}~\lastname{Mu\~noz-Echeverr\'{\i}a}
   \and  \firstname{A.}~\lastname{Paliwal} \inst{\ref{Roma}}
   \and  \firstname{L.}~\lastname{Perotto} \inst{\ref{LPSC}}
   \and  \firstname{G.}~\lastname{Pisano} \inst{\ref{Cardiff}}
   \and  \firstname{E.}~\lastname{Pointecouteau} \inst{\ref{Toulouse}}
   \and  \firstname{N.}~\lastname{Ponthieu} \inst{\ref{IPAG}}
   \and  \firstname{G.~W.}~\lastname{Pratt} \inst{\ref{CEA}}
   \and  \firstname{V.}~\lastname{Rev\'eret} \inst{\ref{CEA}}
   \and  \firstname{A.~J.}~\lastname{Rigby} \inst{\ref{Cardiff}}
   \and  \firstname{A.}~\lastname{Ritacco} \inst{\ref{ENS},\ref{IAS}}
   \and  \firstname{C.}~\lastname{Romero} \inst{\ref{Pennsylvanie}}
   \and  \firstname{H.}~\lastname{Roussel} \inst{\ref{IAP}}
   \and  \firstname{F.}~\lastname{Ruppin} \inst{\ref{MIT}}
   \and  \firstname{K.}~\lastname{Schuster} \inst{\ref{IRAMF}}
   \and  \firstname{S.}~\lastname{Shu} \inst{\ref{Caltech}}
   \and  \firstname{A.}~\lastname{Sievers} \inst{\ref{IRAME}}
   \and  \firstname{C.}~\lastname{Tucker} \inst{\ref{Cardiff}}
   \and  \firstname{G.}~\lastname{Yepes} \inst{\ref{Madrid}}
}

   \institute{
     Univ. Grenoble Alpes, CNRS, LPSC-IN2P3, 53, avenue
des Martyrs, 38000 Grenoble, France
     \label{LPSC}
     \and
     LLR, CNRS, École Polytechnique, Institut Polytechnique de Paris,
Palaiseau, France
     \label{LLR}
     \and
     School of Physics and Astronomy, Cardiff University, Queen’s
Buildings, The Parade, Cardiff, CF24 3AA, UK
     \label{Cardiff}
     \and
     AIM, CEA, CNRS, Universit\'e Paris-Saclay, Universit\'e Paris
Diderot, Sorbonne Paris Cit\'e, 91191 Gif-sur-Yvette, France
     \label{CEA}
     \and
     INAF, IASF-Milano, Via A. Corti 12, 20133 Milano, Italy
     \label{Milano}
     \and
     Institut d'Astrophysique Spatiale (IAS), CNRS, Universit\'e Paris
Sud, Orsay, France
     \label{IAS}
     \and
     Institut N\'eel, CNRS, Universit\'e Grenoble Alpes, France
     \label{Neel}
     \and
     Institut de RadioAstronomie Millim\'etrique (IRAM), Grenoble, France
     \label{IRAMF}
     \and
     Aix Marseille Univ, CNRS, CNES, LAM, Marseille, France
     \label{LAM}
     \and
     Dipartimento di Fisica, Sapienza Universit\`a di Roma, Piazzale
Aldo Moro 5, I-00185 Roma, Italy
     \label{Roma}
     \and
     Univ. Grenoble Alpes, CNRS, IPAG, 38000 Grenoble, France
     \label{IPAG}
     \and
     Centro de Astrobiolog\'ia (CSIC-INTA), Torrej\'on de Ardoz, 28850
Madrid, Spain
     \label{CAB}
     \and
     Instituto de Radioastronom\'ia Milim\'etrica (IRAM), Granada, Spain
     \label{IRAME}
     \and
     LERMA, Observatoire de Paris, PSL Research University, CNRS,
Sorbonne Universit\'e, UPMC, 75014 Paris, France
     \label{LERMA}
     \and
     Univ. de Toulouse, UPS-OMP, CNRS, IRAP, 31028 Toulouse, France
     \label{Toulouse}
     \and
     Laboratoire de Physique de l’\'Ecole Normale Sup\'erieure, ENS, PSL
Research University, CNRS, Sorbonne Universit\'e, Universit\'e de Paris,
75005 Paris, France
     \label{ENS}
     \and
     Department of Physics and Astronomy, University of Pennsylvania,
209 South 33rd Street, Philadelphia, PA, 19104, USA
     \label{Pennsylvanie}
     \and
     Institut d'Astrophysique de Paris, CNRS (UMR7095), 98 bis boulevard
Arago, 75014 Paris, France
     \label{IAP}
     \and
     Kavli Institute for Astrophysics and Space Research, Massachusetts
Institute of Technology, Cambridge, MA 02139, USA
     \label{MIT}
     \and
     School of Earth and Space Exploration and Department of Physics,
Arizona State University, Tempe, AZ 85287, USA
     \label{Arizona}
     \and
     Caltech, Pasadena, CA 91125, USA
     \label{Caltech}
     \and
     Departamento de F\'isica Te\'orica and CIAFF, Facultad de Ciencias,
Modulo 8, Universidad Aut\'anoma de Madrid, 28049 Madrid, Spain
     \label{Madrid}
   }

\abstract{%
  PSZ2 G091.83+26.11 is a massive galaxy cluster with $M_\mathrm{500}= 7.43\times10^{14} M_\odot$ at $ z = 0.822$. This object exhibits a complex morphology with a clear bimodality observed in X-rays. However, it was detected and analysed in the \planck~sample as a single, spherical cluster following a universal profile \cite{2010A&A...517A..92A}. This model can lead to miscalculations of thermodynamical quantities, like the pressure profile. As future multiwavelength cluster experiments will detect more and more objects at higher redshifts (where we expect the fraction of merging objects to be higher), it is crucial to quantify this systematic effect.
In this work, we use high-resolution observations of PSZ2 G091.83+26.11 by the NIKA2 camera to integrate the morphological characteristics of the cluster in our modelling. This is achieved by fitting a two-halo model to 
the SZ image and then by reconstruction of the resulting projected pressure profile. We then compare these results with the spherical assumption.}
\maketitle
\section{Introduction}
\label{intro}
Galaxy clusters are the largest virialized objects in the universe. As such, they can constrain cosmological models, fondamental physics, and astrophysics \cite{Cataneo_2018, 2015PhRvD..92d4009C}. Precisely, the number of clusters per unit of mass and volume, modelled as the halo mass function (historically from \cite{1974ApJ...187..425P}), constitutes a robust cosmological probe, measuring among others the matter density parameter $\om$,  the rms linear fluctuations on scales of 8 $\mathrm{h}^{-1}\mathrm{Mpc}$, $\sig$, the dark energy equation of state parameters, as well as the sum of the mass of the neutrinos \cite{2018A&A...614A..13S, 2019MNRAS.489..401Z}. Cluster counts have been used by many surveys like \planck ~\cite{2016A&A...594A..24P}, \textit {DES} \cite{2020arXiv200211124D}, and \textit{XXL} \cite{2018A&A...620A..10P} to cite a few examples.\\ %The next generation of telescopes will dramatically increase the number of observed galaxy clusters. \euclid \cite{2016MNRAS.459.1764S} and the \rubin~observatory \cite{2019ApJ...873..111I} will both provide cosmological catalogs of $\sim 10^5$ clusters in the optical, \textit{eROSITA} \cite{2012arXiv1209.3114M} will increase our understanding of the low mass range of the cluster regime in for the X-rays, while the \textit{Simons observatory} \cite{Ade_2019}, \textit{CMB-S4} \cite{2016arXiv161002743A} and the proposed mission \textit{PICO} \cite{2019arXiv190807495H} will continuously expand the size of SZ detected cluster catalogs.\\
However, the total mass of dark matter halos hosting clusters is not an observable quantity, and must be inferred from different physical phenomena \cite{2019SSRv..215...25P}, like the thermal Sunyaev-Zel’dovich effect (tSZ, \cite{1970Ap&SS...7....3S}). As a consequence, astrophysical systematic effects are biasing our measurements \cite{Bocquet2015}, and their effects must be integrated in any cosmological analysis. Moreover, given the fact that the size of the catalogs will be increased by several orders of magnitude in the future, effects that are now neglected will play a crucial role \cite{2021A&A...649A..47A, 2021A&A...652A..21F}. \\
The Large program SZ of the NIKA2 experiment (LPSZ, \cite{2020EPJWC.22800017M}) aims at investigating these systematic effects, taking advantage of the spatial resolution and FoV of the NIKA2 camera.  In this work, we focus on the impact of cluster morphology on the reconstruction of observable quantities in the case of the massive cluster PSZ2G091.
\section{The case of PSZ2G091}
\label{sec:psz2G091}
\begin{figure}[h]
\centering
\begin{minipage}{.33\textwidth}
  \centering
  \includegraphics[width=\linewidth]{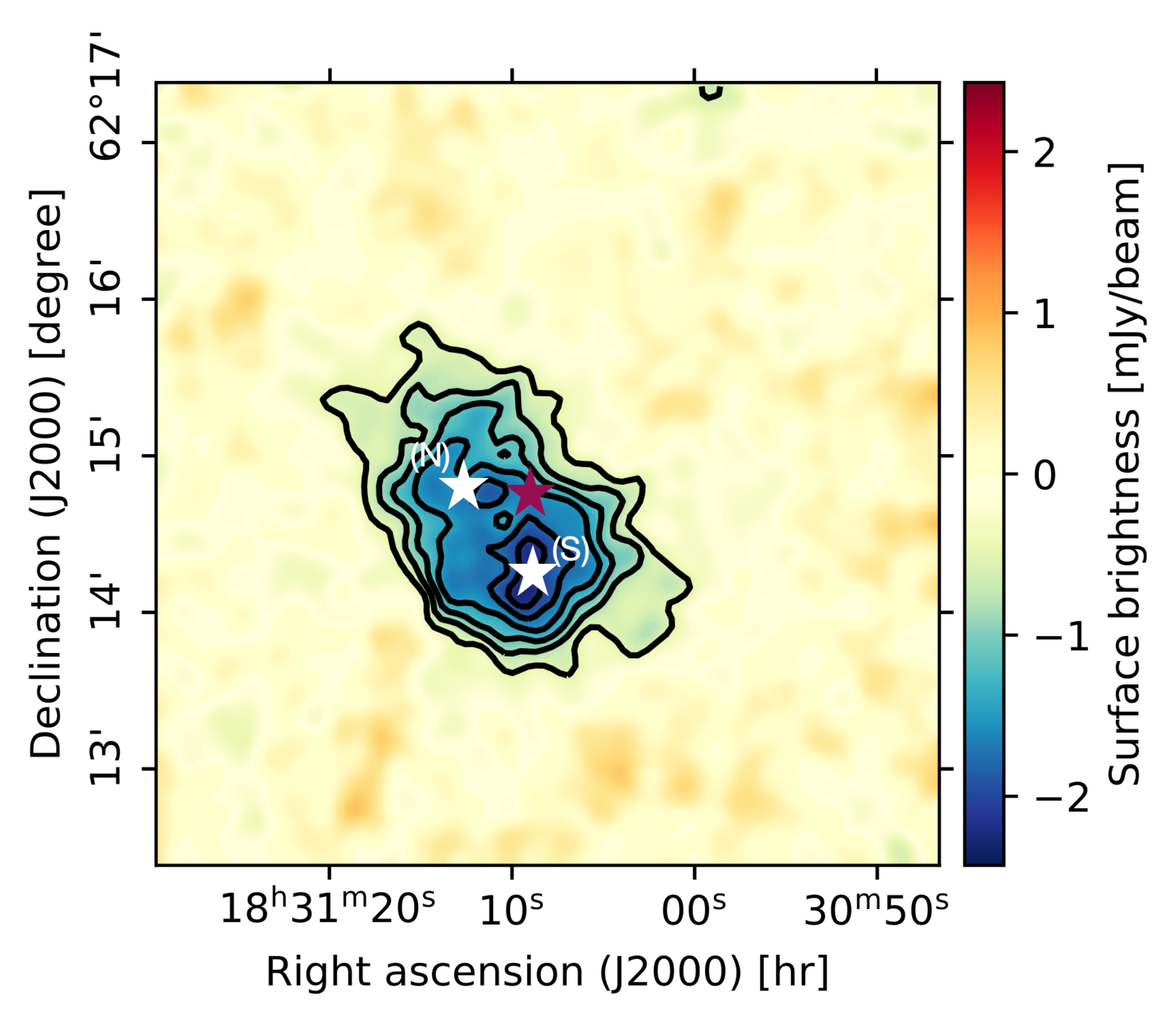}
\end{minipage}%
\begin{minipage}{.33\textwidth}
  \centering
  \includegraphics[width=\linewidth]{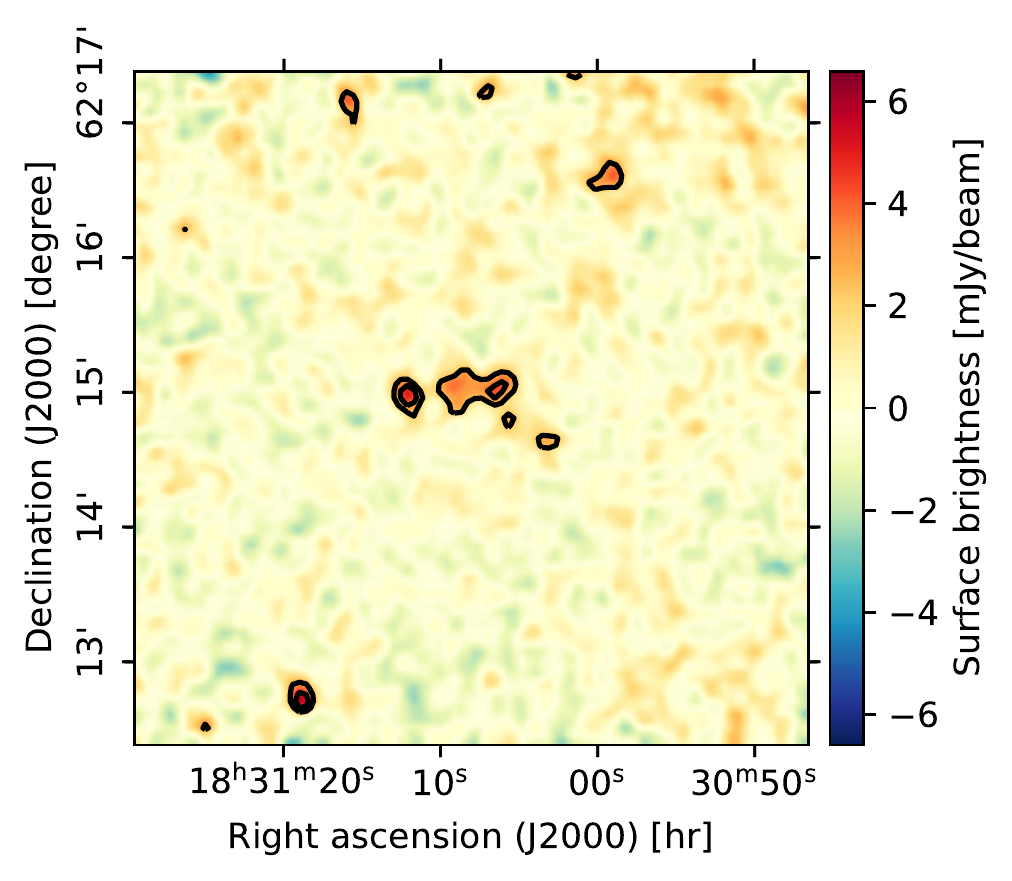}
\end{minipage}
\begin{minipage}{.33\textwidth}
  \centering
  \includegraphics[height=3.8cm,width=\linewidth]{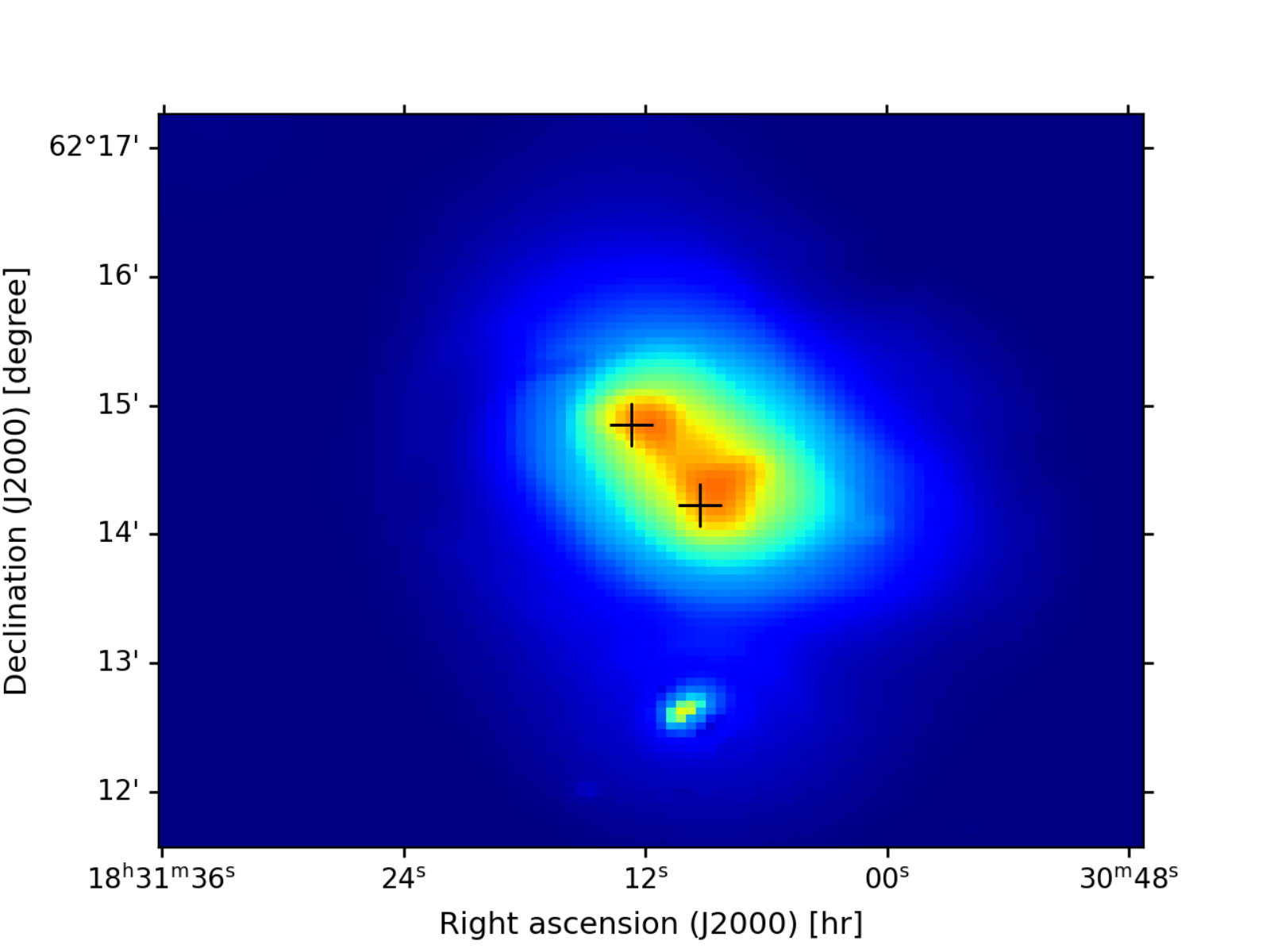}
\end{minipage}

\caption{\textbf{Left}: Surface brightness at 2 mm from NIKA2 observations. The black contours represent the SNR levels starting at  $3\sigma$ increasing by $2\sigma$ at each step. The northern (N) and southern (S) X-ray peaks are shown as white stars, the X-ray centroid being in purple. \textbf{Middle}: Surface brightness at 1 mm. The signal exhibits several point sources. Their treatment is explained in section \ref{sec:point_sources}. \textbf{Right}: X-ray surface brightness of PSZ2G091 from \textit{XMM-Newton}. The cluster exhibits two peaks at the position $(\alpha_\mathrm{N},\delta_\mathrm{N})=(277.80,62.24)$, and $(\alpha_\mathrm{S},\delta_\mathrm{S})=(277.79, 62.237)$, shown with the black crosses.}
\label{fig:nika2_2mm}
\end{figure}
%
%PSZ2 G091.83+26.11 is part of the SZ Large Program of the NIKA2 experiment \cite{2020EPJWC.22800017M}. It was, before these observations, detected by \planck \cite{2016A&A...594A..27P}, and was also observed by XMM-\textit{Newton}. The characteristics
%of this cluster are shown in table \ref{tab:cluster_car}.
\begin{table}[h]
\centering
\caption{Characteristics of PSZ2 G091.83+26.11 as described in \cite{2016A&A...594A..27P}, with the observation length and SNR at SZ peak.}
\label{tab:cluster_car}       % Give a unique label
% For LaTeX tables you can use
\begin{tabular}{lllll}
\hline \hline
$z$ & $M_\mathrm{500}$ & $\theta_\mathrm{500}$ & $t_\mathrm{obs}/t_\mathrm{LPSZ}$& tSZ decrement peak  \\\hline \hline
0.822 & $7.43\times 10^{14}M_\odot$ & $2.2~\mathrm{arcmin}$ & 2.5h/2.5h=1 & 14.9$\sigma$ \\\hline
\end{tabular}
% Or use
%\vspace*{5cm}  % with the correct table height
\end{table}
\noindent As part of the LPSZ, PSZ2G091 was observed in October, 2017, with an average elevation of $58.5^\circ$ and an average atmospheric opacity at 225 GHz of 0.243. These conditions are standard for observations at the IRAM 30 m telescope, at this season.\\
The data from NIKA2 were reduced following the pipeline described in \cite{2020A&A...637A..71P}. In figure \ref{fig:nika2_2mm}, we show the results of the data reduction at 1 and 2 mm. The cluster can clearly be seen in the 2 mm image, and it is clearly elongated in the NW-SW direction. There is a clear departure from sphericity, and a hint of bimodality later confirmed in the X-ray surface brightness. Conversely, in the 1 mm map, we only detect point sources, since the SZ signal is completely dominated by the noise as expected. The peaks in the X-ray map are in good agreement with the one observed in the NIKA2 2 mm map. This would imply the presence of two well-defined sub-halos in the first stages of a major merger.
\section{Point sources treatment}
\label{sec:point_sources}
The 1 mm map of PSZ2G091 exhibits 7 point sources (see figure \ref{fig:nika2_2mm}). Their fluxes must be inferred at 2 mm, in order to fit the ICM signal. If external information on these sources is available, we can usually use SED models to infer a PDF of the flux at 2 mm for each source, and use it as a prior for the model reconstruction. However, in this case, no additional data was available. The point sources observed at this wavelength are either millimetric sources, or radio objects. For the latter, the flux expected at 2 mm is higher than the one seen in the 1 mm map. Given the value of the SZ signal in the 2 mm map, we were consequently able to identify the sources as millimetric sources.Thus we decided to used flat priors for the fluxes of the sources at 2 mm, with a conservative upper limit of $F_{\mathrm{2,upper}}=F_1/3$. The fluxes of the point sources were then fitted jointly with the parameters of the profiles, using these flat priors.
\section{Imaging analysis}

\subsection{Single spherical model}
\label{sec:single_model}
As a first test case, we consider a single spherical halo centred on the X-ray centroid coordinates. A forward modelling approach is incorporated in an MCMC sampling framework to fit the parameters of the pressure profile, as well as the point sources, as described in section \ref{sec:point_sources}, with the collaboration software PANCO2. Instead of considering a generalised Navarro-Frenk-and-White profile \cite{2007ApJ...668....1N}, we use a power law model, where the pressure is divided into 6 bins, equally distributed in \textit{log}-space. Each one of these bins follow the identity
\begin{equation}
\label{eq:power_law}
P(r)=P_i(r/r_i)^{-\alpha_i}.
\end{equation}
The top row of figure \ref{fig:spherical_model} shows the results of the fitting procedure. Although returning reasonable residuals, the spherical symmetry clearly does not encapsulate the bimodal nature of the cluster. It is then required to improve our modelling.

\subsection{Two-halo model}
\label{sec:bimodal_cluster}
\begin{figure}
\centering
\begin{subfigure}[b]{\textwidth}
   \includegraphics[width=1\linewidth]{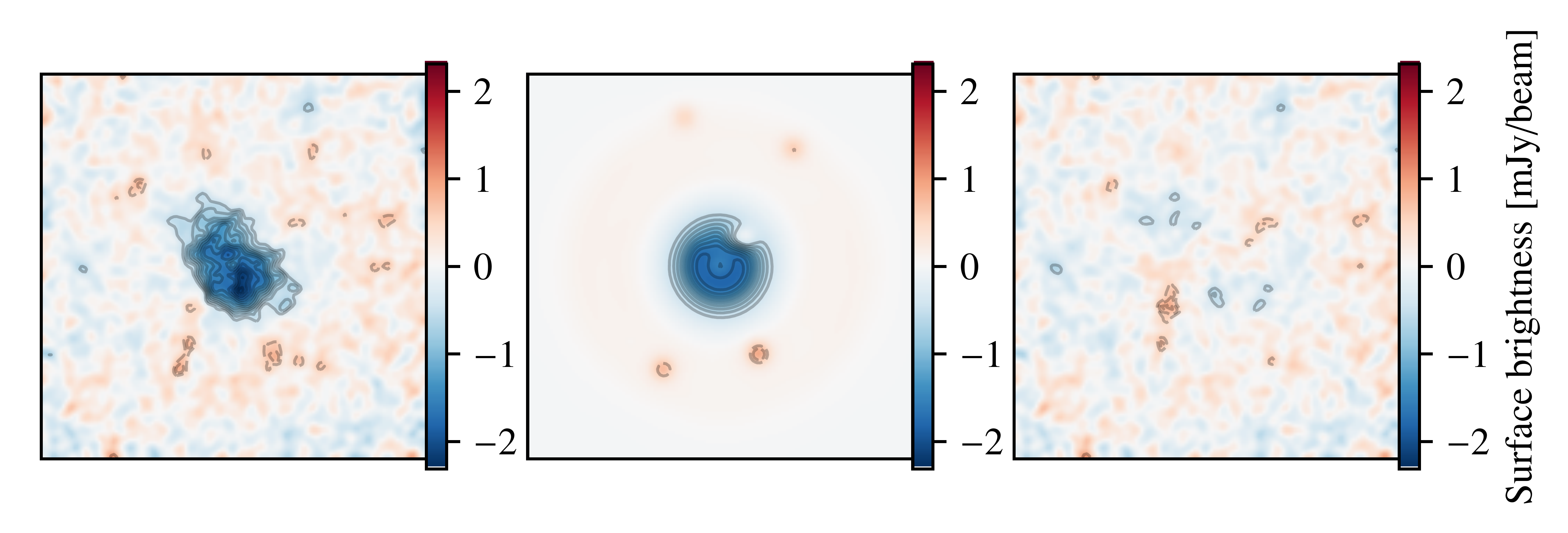}
   \end{subfigure}

\begin{subfigure}[b]{\textwidth}
   \includegraphics[width=1\linewidth]{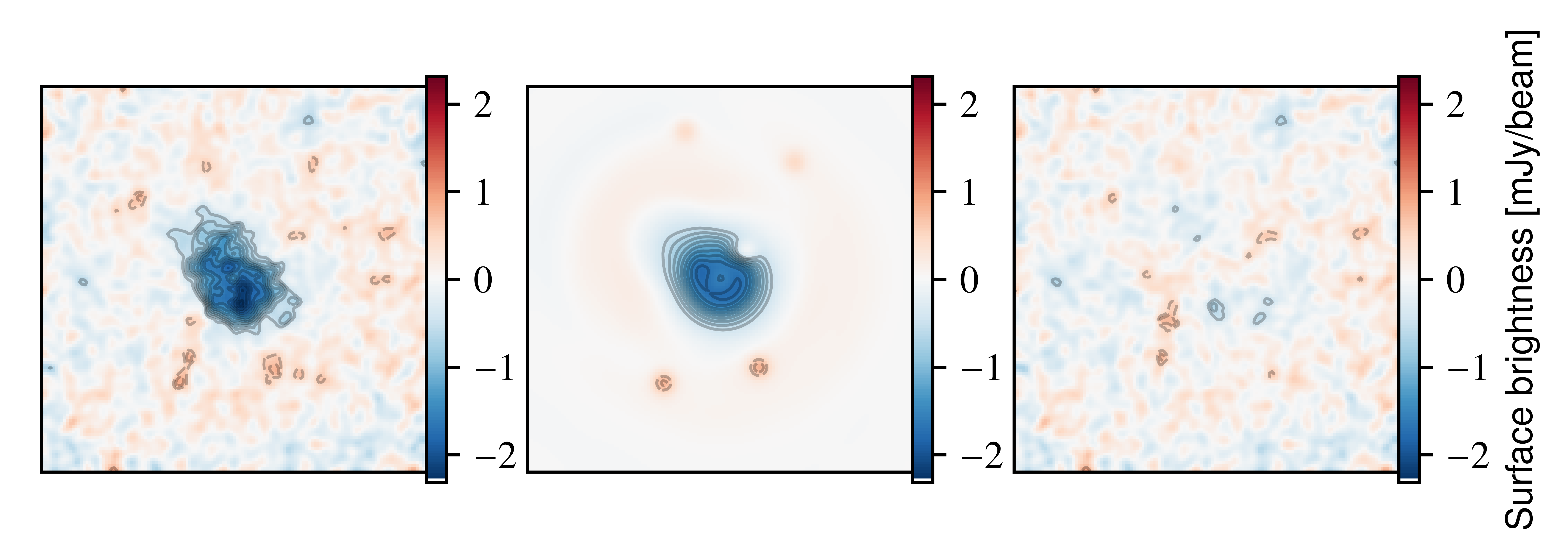}
\end{subfigure}
\caption{Single spherical model fit of the NIKA2 2 mm map for PSZ2G091 centered on the X-ray centroid coordinates (\textbf{top}), and 2-halo model (\textbf{bottom}). From \textbf{left} to \textbf{right}, we display the data, the single halo model, with the point sources treated as described in section \ref{sec:point_sources}, and the residuals. The maps are given in a $5'\times5'$ area, and for display purposes, each map is smoothed with a gaussian kernel. The contours are showing the SNR level sets, starting at $3\sigma$ and increasing with a step of $1\sigma$.}
\label{fig:spherical_model}
\end{figure}

%
%\begin{figure}[h]
%\begin{center}
%\includegraphics[scale=0.25]{Figures/x_ray_brightness.pdf}      
%\caption{X-ray surface brightness of PSZ2G091 from \textit{XMM-Newton}. The cluster exhibits two peaks at the position $(\alpha_\mathrm{N},\delta_\mathrm{N})=(277.80,62.24)$, and $(\alpha_\mathrm{S},\delta_\mathrm{S})=(277.79, 62.237)$.}
%\label{fig:x_ray_b} 
%\end{center}
%\end{figure}
%
%
\noindent Instead of considering a single pressure profile, we jointly fit two halos at the positions of the X-ray peaks. %
The top row of figure \ref{fig:spherical_model} shows the results of the MCMC fitting.
The results of the fits are shown on the bottom row of figure \ref{fig:spherical_model}. It is clear that the two halo model yields a more realistic representation of the dynamical state of the cluster. Additionally, the residuals are slightly improved in the region of the northern subhalo. Of course, due to the non-spherical nature of this cluster, it is not possible to consider a radial pressure profile. However, in section \ref{sec:thermo}, we describe how we recover an average radial profile for the two-halo model.
\section{Pressure profile reconstruction}
\label{sec:thermo}
Our model takes into account the specific morphology of this cluster. However, the thermodynamical profiles are usually considered with the goal of reconstructing a mass profile, using the hydrostatic equilibrium equation. This requires the presence of a 1D pressure profile,
\begin{figure}[h]
\begin{center}
\includegraphics[scale=0.7]{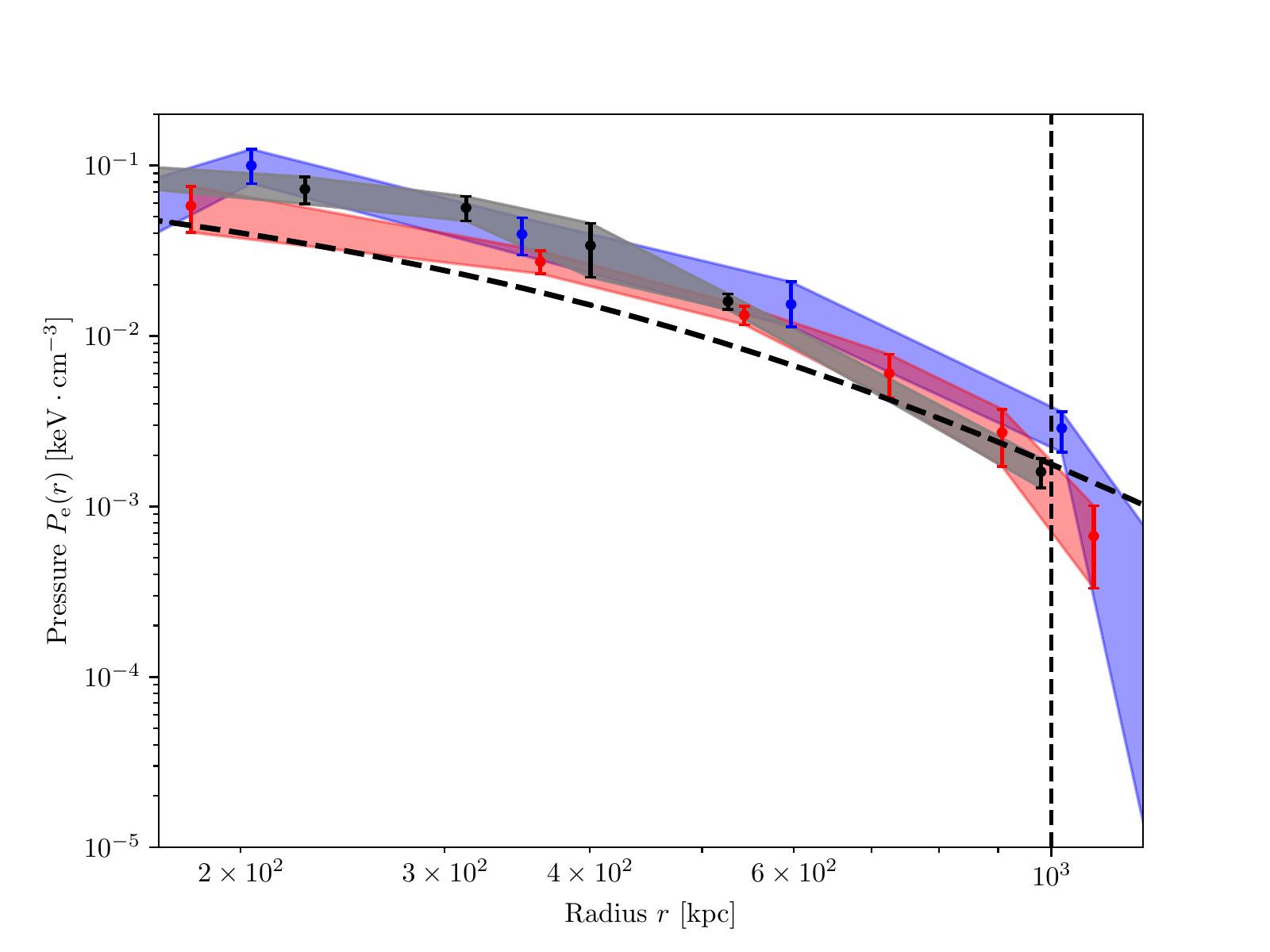}
\caption{In \textcolor{red}{red}, composite pressure profile obtained by combining the northern (N) and southern (S) profiles, as described in equation \ref{eq:mean_prof}. In \textcolor{blue}{blue}, profile obtained with the single spherical model. In black, we represent the contours obtained from XMM X-ray data only.  The  black dashed line represents the universal profile from \cite{2013A&A...550A.131P}. The vertical dashed line represents the radius $R_\mathrm{500}$ obtained from X-ray data assuming spherical symmetry. Although the single spherical model seems to slightly overestimate the pressure, both profile look compatible for the considered radius range.}%The radius $R_S$ correspond to the radius $R_{500}$ obtained for a spherically symmetric profile. The black line is the mean pressure profile. The blue error bars correspond to the 68\% confidence contours. }
\label{fig:mean_pres}       % Give a unique label
\end{center}
\end{figure}
when we previously fitted a 2D map. Thus, we use the following procedure to recover an average 1D profile. We consider both the profiles of the two halos fitted in section \ref{sec:bimodal_cluster}. With the assumption of the two sub-halos lying in the same plane perpendicular to the line of sight, we recover a mean radial pressure profile by integrating both the profiles in annuli centered around the X-ray centroid coordinates. In this framework, the pressure in the $i$-th bin reads:
\begin{equation}
\label{eq:mean_prof}
P_i=\left(\int_{r_i}^{r_{i+1}}(P_N(r)+P_S(r))2\pi r \mathrm{d}r\right)/\left(\pi(r_{i+1}^2-r_i^2)\right),
\end{equation}
where $P_i$ is the value of the pressure, $P_N$ and $P_S$ are respectively the profiles of the northern and southern subhalos, $r_i$ and $r_{i+1}$ being the inner and outer radii of the annulus. This allows us to reconstruct the profile available in figure \ref{fig:mean_pres}. The integration is performed up to the radius $R_\mathrm{500}$ that is found with the X-ray data. The 1D and average 2D models seem to be slightly compatible,  which tends to show that even in the case of an highly disturbed cluster, the spherical model still yields robust results.
%For this specific analysis, the radius $R_\mathrm{500}$ is not defined.  Instead, we define the specific radius $R_\mathrm{S}$, which is equal to the radius $R_{500}$ for the case of a single spherical model obtained from X-rays.
%\begin{figure}[h]
%\begin{center}
%\includegraphics[scale=0.4]{Figures/mean_density.pdf}
%\caption{blabla}
%\label{fig:mean_dens}       % Give a unique label
%\end{center}
%\end{figure}
%
\section{Conclusions}
\label{sec:conc}
This analysis shows the challenges related to the complexity of cluster morphologies. In this work, we showed that for the case of the highly disturbed cluster PSZ2 G091, taking into account the merging state of the cluster yields results compatible with the spherical profile. This is a promising result, as the pressure profile impacts the $Y_\mathrm{500}-M$ relation. To complete this analysis, we plan to perform a full thermodynamical analysis of this cluster, recovering the 2D maps of physical quantities like the temperature and the entropy. This a precondition to assess the impact of the morphology of this cluster on its full mass reconstruction, and generally on cosmological inference using clusters.  %A full multiwavelength morphological analysis is required before moving on to a more precise mass reconstruction model.
%As cluster surveys will go deeper in term of redshift, we expect the fraction of merging objects to increase. The analysis of the systematics related to these objects will be a major issue for future cluster surveys.

\section*{Acknowledgements}
\small{We would like to thank the IRAM staff for their support during the campaigns. The NIKA2 dilution cryostat has been designed and built at the Institut N\'eel. In particular, we acknowledge the crucial contribution of the Cryogenics Group, and in particular Gregory Garde, Henri Rodenas, Jean Paul Leggeri, Philippe Camus. This work has been partially funded by the Foundation Nanoscience Grenoble and the LabEx FOCUS ANR-11-LABX-0013. This work is supported by the French National Research Agency under the contracts "MKIDS", "NIKA" and ANR-15-CE31-0017 and in the framework of the "Investissements d’avenir” program (ANR-15-IDEX-02). This work has benefited from the support of the European Research Council Advanced Grant ORISTARS under the European Union's Seventh Framework Programme (Grant Agreement no. 291294). F.R. acknowledges financial supports provided by NASA through SAO Award Number SV2-82023 issued by the Chandra X-Ray Observatory Center, which is operated by the Smithsonian Astrophysical Observatory for and on behalf of NASA under contract NAS8-03060.}

%For tables use syntax in table~\ref{tab-1}.
%\begin{table}
%\centering
%\label{tab-1}       % Give a unique label
% For LaTeX tables you can use
%\begin{tabular}{lll}
%\hline
%first & second & third  \\\hline
%number & number & number \\\hline
%\end{tabular}
% Or use
%\vspace*{5cm}  % with the correct table height
%\end{table}
%
% BibTeX or Biber users please use (the style is already called in the class, ensure that the "woc.bst" style is in your local directory)
% \bibliography{name or your bibliography database}
\bibliography{references}
%\begin{thebibliography}{}
%Journal Author, Journal \textbf{Volume}, page numbers (year)
%Book Author, \textit{Book title} (Publisher, place, year) page numbers
% etc
%\end{thebibliography}

\end{document}